\documentclass[aip,twocolumn,letterpaper]{revtex4}

\usepackage{epsfig}

\usepackage{graphics}
\usepackage{epsfig}

\usepackage[margin=1.0in]{geometry}

\begin{document}
\title{Efficient analysis of magnetic field line behavior in toroidal plasmas }
\author{Allen H Boozer}
\affiliation{Columbia University, New York, NY  10027 \linebreak ahb17@columbia.edu}

\begin{abstract} 
The confinement of plasmas in tokamaks and stellarators depends on magnetic field lines lying in nested toroidal surfaces.  The transition near the plasma edge away from the lines lying in magnetic surfaces defines properties of divertors.  The transition in time defines properties of disruptions.   Divertor design and disruption analyses require a detailed understanding of these transitions.  The use of a Fourier transform coupled with a Gaussian window function allows far more information to be extracted about these transitions using far shorter field line integrations than can be obtained using traditional methods based on Poincar\'e plots.  The physics of divertors and disruptions is reviewed to clarify why the type of information that can be gained from more efficient methods of analysis is of central importance to the fusion program based on magnetic confinement.

 \end{abstract}

\date{\today} 
\maketitle

%%%%%%%%%%%%%%%%%%%%%%%%%%%%%%%%%%%%%%%%%%%%%%%%%%%%%%%%%%

\section{Introduction}

Magnetic field lines that lie on surfaces are the central concept to the confinement of plasmas in tokamaks and stellarators.  The transition in either space or time away from nested magnetic surfaces must be understood to understand divertors, disruptions, and runaway electron effects.  Section \ref{Importance} explains the practical importance of having an efficient method of studying such transitions, and Section \ref{sec:Gaussian} explains how Fourier transforms coupled with Gaussian window functions provides  such a method. % Reviews of the physics and mathematical considerations that make an understanding both subtle and important are given in Section \ref{Divertors} for divertors and Section \ref{Disruptions and runaway} for disruptions and runaways.
%%%%

The simplest example of nested confining surfaces becoming leaky is in a 1984 paper \cite{turnstiles} by  MacKay, Meiss, and Percival, which illustrates basic concepts using the standard map.  The Cantorus, which they illustrate, is the type of leaky surface that it is important to able to efficiently construct.  They called the leaks themselves turnstiles because the flux of lines leaking out of a cantorus of the standard map must equal the flux of lines leaking in, as they must in any Hamiltonian system.  Magnetic field lines are a Hamiltonian system \cite{Boozer:RMP}.

As the standard map example of MacKay et al illustrates, the transition in space or time away from nested confining surfaces is related to a general problem in Hamiltonian mechanics and related fields, leaking chaotic systems.  This topic was the subject of a 2013 article by Altmann, Portela, and T\'el in the Reviews of Modern Physics \cite{Leaky systems:2013}.  Section V.B,  the ``Decay of the survival probability in open systems," points out that the presence of ``sticky regions" implies the asymptotic decay rate of the number of confined trajectories switches from exponential to a power law.  A weak leak through a surface makes that surface sticky for trajectories.  The 2010 article by G. Contopoulos and M. Harsoula \cite{Stickiness:2010}, contains discussions and figures that are relevant to non-resonant divertors as well as the breakup of magnetic surfaces in disruptions.  Stickiness is important for a strongly perturbed single-null axisymetric divertor as shown in the 2019 paper by Martins, Roberto, and Caldas \cite{Stickiness:2014}.  In 1983, Charles Karney \cite{Karney:1983} introduced the concept of stickiness from his studies of plasma problems involving Hamiltonian trajectories crossing phase space regions with islands.  George Zaslavsky \cite{Zaslavsky:2002} wrote a well-known review of the theory of chaos in 2002.

%%%%%%

The primary topic of this paper is the use of a Fourier transform coupled with a Gaussian window function, Section \ref{sec:Gaussian}, to determine through a relatively short field line integration whether a particular magnetic field line lies on an analytic or a rough magnetic surface or chaotically fills a volume.  The basic concept used in Section \ref{sec:Gaussian} is simple and was used in 1982 by Boozer \cite{Boozer:1982} to give the first general method of numerically determining magnetic coordinates.  

The Gaussian window function was introduced by Dennis Gabor in 1946 \cite{Gabor:1946} for the compression of data in frequency for efficient transmission.  A web search will demonstrate that there are many window functions.  The optimality of the Gaussian window function was the subject of a 1991 paper by Augustus Janssen \cite{Gaussian-window:1991}.

Using $(R,\varphi,Z)$ cylindrical coordinates, a line that lies on a magnetic surface can be described by the functions $R(\Theta,\varphi)$ and $Z(\Theta,\varphi)$ that give the surface on which the line lies.  The poloidal angle $\Theta$ can be taken to have the form $\Theta = N_p\iota_p\varphi$ along the line with $N_p$ the number of toroidal periods and $\iota_p$ a constant, the rotational transform per period \cite{Boozer:RMP}.  An axisymmetric tokamak has one period but stellarators have a number, for example the W7-X stellarator has five.  The angle $\zeta$,
\begin{eqnarray}
\zeta &\equiv& N_p \varphi \mbox{    so   } \\
\Theta &=& \iota_p \zeta, \label{mag-Theta}
\end{eqnarray}
allows one to take full advantage of the periodicity, and will be used as the toroidal angle.

Expanded in Fourier series in $\Theta$ and $\zeta$, the Fourier coefficients $Z_{mn}$ and $R_{mn}$ converge exponentially when $R(\Theta,\zeta)$ and $Z(\Theta,\zeta)$ are analytic functions.  That is positive constants $c_1$ and $c_2$ exist such that $c_1 e^{-c_2 \sqrt{m^2+n^2}}> |Z_{mn}|$ for all $m$ and $n$.  The surface has increasing roughness the smaller the power $p$ required to ensure $c/ \big(\sqrt{m^2+n^2}\hspace{0.03in} \big)^p> |Z_{mn}|$ for all $m$ and $n$.  When the line is chaotic, the Fourier coefficients have an intrinsic width when determined using the Gaussian window function method, as discussed in Section \ref{sec:Gaussian}.

Section \ref{sec:applications} explains analyses that can be carried out using the functions $R(\Theta,\zeta)$ and $Z(\Theta,\zeta)$ that define a surface.   Expressions are given (1) for the roughness $\mathcal{R}$ of the magnetic surface, (2) the leakage $\psi_\ell$ of magnetic flux through the surface and the collimation $\mathcal{K}$ of the leaked flux, and (3) the magnetic field line Hamiltonian $\Psi_p(\Psi,\Theta,\zeta)$.  

A rough surface is extremely sensitive to breakup through the diffusion of magnetic field lines by resistivity, $-\vec{\nabla}\times (\eta \vec{j}/\mu_0)\approx (\eta/\mu_0)\nabla^2\vec{B}$. 

In transition regions between magnetic field lines lying on surfaces and being chaotic, the surfaces defined by $R(\Theta,\zeta)$ and $Z(\Theta,\zeta)$ leak magnetic flux in collimated tubes, called turnstiles.   Magnetic field lines are chaotic in a spatial volume when each line has neighboring lines that exponentially separate from it with distance $\ell$ along the line.

The field line Hamiltonian $\Psi_p(\Psi,\Theta,\zeta)$ is an analytic function even as the field line trajectories transition from lying on analytic surfaces to being chaotic.  The Hamiltonian takes a simple form $\Psi_p(\Psi)$, when analytic magnetic surfaces exist  \cite{Boozer:RMP}.

Section \ref{chaos-cyl} gives a method for determining the separation of neighboring field lines from a given line by integrating the trajectory of that line plus a pair linear ordinary differential equations.  When the number of e-folds of separation in a region is larger than the natural logarithm of the resistive diffusion timescale divided by the evolution timescale, large scale reconnection will occur.

%%%%%%%%%%%%%%%%%%%%%%%%%%%%

%%%%%%%%%%%%%%

\section{Gaussian method of magnetic surface Identification \label{sec:Gaussian}}

An understanding of the multiple issues involving divertors and disruptions require detailed knowledge of the dependence of the shapes, $R(\Theta,\zeta)$ and $Z(\Theta,\zeta)$, of magnetic surfaces on space and time.  These can be found by integrating various field lines at fixed instants of time.  Each field line, $\vec{X}(\zeta) = R(\zeta)\hat{R}(\zeta) + Z(\zeta)\hat{\zeta}$, is determined by integrating the equation
\begin{eqnarray}
\frac{d\vec{X}}{d\zeta} &=& \frac{dR}{d\zeta}\hat{R}+\frac{R}{N_p}\hat{\zeta} +  \frac{dZ}{d\zeta}\hat{Z}; \\
& = & \frac{\vec{B}(\vec{X})}{\vec{B}\cdot\vec{\nabla}\zeta} \hspace{0.2in} \mbox{   so  }\\
\frac{dR}{d\zeta} &=& \frac{R}{N_p} \frac{\vec{B}(R,\zeta,Z)\cdot \hat{R}}{\vec{B}\cdot\hat{\zeta}}; \\
\frac{dZ}{d\zeta} &=& \frac{R}{N_p} \frac{\vec{B}(R,\zeta,Z)\cdot \hat{Z}}{\vec{B}\cdot\hat{\zeta}}.
\end{eqnarray}
The equations $d\hat{R}/d\zeta=\hat{\zeta}/N_p$, $\vec{\nabla}\zeta = (N_p/R)\hat{\zeta}$ and $\hat{\zeta}=\hat{\varphi}$ were used.

Poincar\'e plots are the standard way to visualize the solution $R(\zeta),Z(\zeta)$.  A point is placed on the $R-Z$ plane each time the field line increases it position in $\zeta$ by a multiple of $2\pi$.  A Poincar\'e plot provides far less information and requires a far longer field line integration than determining Fourier coefficients of $R(\Theta,\zeta)$ and $Z(\Theta,\zeta)$.

To determine the Fourier coefficients of $R(\Theta,\zeta)$ and $Z(\Theta,\zeta)$ use Equation (\ref{mag-Theta}), $\Theta=\iota_p\zeta$, to write
\begin{eqnarray}
Z(\Theta,\zeta) &=& \sum_{mn} Z_{mn} \cos(m\Theta - n\zeta) \hspace{0.2in}\mbox{as }\\
Z(\zeta) &=& \sum_{mn}Z_{mn} \cos(\omega_{mn}\zeta) \hspace{0.2in}\mbox{with   } \\
 \omega_{mn} &\equiv& m\iota-n.
\end{eqnarray}
For simplicity, up-down (stellarator) symmetry in the $\zeta=0$ plane is assumed.

The coefficients $Z_{mn}$ can be found by Fourier transforming $Z(\zeta)$ using a Gaussian window function $G(\zeta)$ of width $\lambda$ in $\zeta$. 
\begin{equation}
G(\zeta)\equiv \frac{1}{\lambda\sqrt{\pi/2}} e^{-\frac{1}{2}\left(\frac{\zeta}{\lambda}\right)^2}.  \label{Gaussian}\
\end{equation}
The Gaussian window function obtains much of its importance from the integral
\begin{equation}
\int_0^\infty G(\zeta) \cos(\Omega\zeta) d\zeta = e^{-\frac{1}{2}(\lambda \Omega)^2}. \label{Gaussian integral}
\end{equation} 

The Fourier transform of $Z(\zeta)$ using a Gaussian window function is 
\begin{eqnarray}
Z_f(\omega) &\equiv& 2 \int_0^\infty G(\zeta) \cos(\omega\zeta) Z(\zeta) d\zeta \label{fourier}\\
&=& 2\sum_{mn} Z_{mn} \int_0^\infty G(\zeta) \cos(\omega\zeta)\cos(\omega_{mn}\zeta) d\zeta \nonumber \\
&=&\sum_{mn}  Z_{mn}\int_0^\infty G(\zeta) \Big\{\cos\Big((\omega-\omega_{mn})\zeta\Big) \nonumber\\
&&\hspace{0.2in} + \cos\Big((\omega+\omega_{mn})\zeta\Big)\Big\}d\zeta \nonumber \\
&=& \sum_{mn}  Z_{mn} \Big\{e^{-\frac{\lambda^2}{2}(\omega-\omega_{mn})^2} \nonumber \\ && \hspace{0.2in} +e^{-\frac{\lambda^2}{2}(\omega+\omega_{mn})^2} \Big\}. \label{transform}
\end{eqnarray} 
When the window is sufficiently wide, $\lambda|\omega_{mn}| >> 1$, then as $\omega \rightarrow \omega_{mn}$
\begin{equation}
Z_f(\omega) \rightarrow Z_{mn} e^{-\frac{\lambda^2}{2}(\omega-\omega_{mn})^2}.
\end{equation}

When a fast Fourier Transform is used $Z_f(\omega)$ will have Gaussian peaks with a standard deviation $1/\lambda$ at each non-zero Fourier term $Z_{mn}$.  The amplitude of the peak gives $Z_{mn}$ and its location at $\omega_{mn}$ gives the transform and the identification of $m$ and $n$.  

Once the transform per period $\iota_p$ is determined, the positions and the amplitudes of Fourier coefficients $Z_{mn}$ and $R_{mn}$ can be quickly found. 
 
When the field line is chaotic, the Gaussian peaks have an additional standard deviation due to their chaotic spreading. When the chaos is small, the standard deviation increases exponentially with the width of the $\zeta$-window, $\lambda$.  Whether the field line is chaotic or not is determined by making the Gaussian width $\lambda$ greater, appropriately extending the integration range, and observing whether the Gaussian peaks become narrower or broader.

Section  \ref{chaos-cyl} gives a direct method of determining whether a particular field line is chaotic, which means have lines in its neighborhood that separate from it exponentially as $\zeta$ increases.  

Appendix \ref{Discrete} explains the use of discrete sums to evaluate the integral of Equation (\ref{transform}), which gives insight into the convergence properties.

%================

\section{Applications of $R(\Theta,\zeta)$ and $Z(\Theta,\zeta)$ \label{sec:applications} }

\subsection{Roughness of magnetic surfaces}

The roughness of magnetic surfaces can be quantified by comparing the circumference $\mathcal{C}(\zeta)$ of a surface at a value of $\zeta$ to its average area $\mathcal{A}$, which is defined by its enclosed toroidal flux $\psi$ divided by the average magnetic field strength $<B>\equiv\int_0^\infty G(\zeta) B(\zeta) d\zeta$ along the magnetic field line.  For a circle, $\mathcal{C}^2=4\pi\mathcal{A}$. The roughness gives the enhancement over this value
\begin{eqnarray}
\mathcal{R}(\zeta)&\equiv& \frac{\mathcal{C}^2}{4\pi\mathcal{A}} \\
\mathcal{C}(\zeta) & =&  \int_0^{2\pi} \sqrt{\left(\frac{\partial R}{\partial \Theta}\right)^2+\left(\frac{\partial Z}{\partial \Theta}\right)^2}d\Theta. \hspace{0.2in}
\end{eqnarray}

A rough surface is extremely sensitive to breakup through the diffusion of magnetic field lines by resistivity, $-\vec{\nabla}\times(\eta\vec{j}/\mu_0)$.

%%%%%%%%%%%%%%%%%%%%%%%%%%%%%%%%%%%%%%%%

\subsection{The unit normal to a $\vec{X}_s(\Theta,\varphi)$ surface}

The unit normal field to the $\vec{X}_s(\Theta,\zeta)$ surface is 
\begin{eqnarray}
\hat{n}(\Theta,\zeta)  &\equiv& \frac{\frac{\partial \vec{X}_s}{\partial\Theta} \times \frac{\partial \vec{X}_s}{\partial\zeta}}{\left| \frac{\partial \vec{X}_s}{\partial\Theta} \times \frac{\partial \vec{X}_s}{\partial\zeta} \right| }. \label{unit norm}
\end{eqnarray}
Note the unit vector $\hat{R}(\zeta)$ has the derivative $d\hat{R}/d\zeta = \hat{\zeta}/N_p$.

The extent to which a surface  $\vec{X}_s(\Theta,\zeta)$ deviates from a magnetic surface is measured by the magnetic flux $\psi_\ell$ that leaks through it in the two directions.  $\vec{\nabla}\cdot\vec{B}=0$ ensures the inward and outward leaking fluxes must be equal.  The expression for the leakage flux is
\begin{eqnarray}
\psi_\ell \equiv \oint \left| \vec{B}\cdot \left(\frac{\partial \vec{X}_s}{\partial\Theta} \times \frac{\partial \vec{X}_s}{\partial\zeta} \right) \right| d\Theta d\varphi.
\end{eqnarray}

The flux leakage can be non-zero for two reasons: (1) The surface $\vec{X}_s(\Theta,\zeta)$ has not been found with sufficient accuracy.  (2) A precise magnetic surface does not exist.

When a precise magnetic surface does not exist as in a Cantorus, the flux leakage can be concentrated in certain regions on the surface.  These regions called turnstiles are penetrated by collimated flux tubes.  This collimation is quantified by
\begin{equation}
\mathcal{K} \equiv \frac{ (2\pi)^2\oint \left| \vec{B}\cdot \left(\frac{\partial \vec{X}_s}{\partial\Theta} \times \frac{\partial \vec{X}_s}{\partial\zeta} \right) \right|^2 d\Theta d\varphi }{ \psi_\ell^2  }.
\end{equation} 
A large collimation $\mathcal{K}$ is an expected on a Cantorus.

Where a magnetic field is transitioning from nested magnetic surfaces to a large region of chaotic lines, separate annuli of chaos are separated by surfaces on which the transform $\iota_p$ is difficult to approximate by a rational number, the ratio of two integers $N/M$.  That is $|\iota_p-N/M|<\epsilon$ requires very large integers for a small $\epsilon$.  These are the most robust surfaces to breaking, and they break with collimated flux tubes expanding the region covered by  each chaotic field line.   These barriers to chaos can be located by finding the leakage flux and its collimation on surfaces determined by integrating field lines.  As discussed in Section \ref{chaos-cyl}, the field line integrations can determine whether a particular field line has neighboring lines that have a separation from it that increases exponentially with $\zeta$, which is the definition of chaos, and where in space that increase is particularly large or small.

%%%%%%%%%%%%%%%%%%%%%%%%%%

\subsection{Enclosed toroidal flux}

%%%%%%%%%%%%%

When the Fourier amplitudes have well defined magnitudes of $R(\Theta,\zeta)$ and $Z(\Theta,\zeta)$,
\begin{equation}
\vec{X}_s(\Theta,\zeta) = R(\Theta,\zeta) \hat{R}(\zeta) + Z(\Theta,\zeta) \hat{Z}
\end{equation}
defines a surface, which closely approximates a magnetic surface.  The toroidal magnetic flux $\psi$ that is enclosed by that surface is an integral over $\vec{B}\cdot\hat{\zeta}$ with respect $R$ and $Z$ at any fixed value of $\zeta$.  The independence of the toroidal flux from the toroidal angle follows from $\vec{\nabla}\cdot\vec{B}=0$ when $\vec{X}_s(\Theta,\zeta)$ is a magnetic surface.

To calculate $\psi$, let $R(\Theta,\zeta) = R_0(\zeta) + \tilde{R}(\Theta,\zeta)$ and $Z(\Theta,\zeta) = Z_0(\zeta) + \tilde{Z}(\Theta,\zeta)$, where $R_0(\zeta)$ and $Z_0(\zeta)$ form a closed curve, such as the magnetic axis, that is enclosed by the magnetic surface.  Then, an arbitrary location $(R_a,Z_a)$ inside the magnetic surface can be written as $R_a = R_0 + r \tilde{R}(\Theta,\zeta)$ and $Z_a = Z_0 + z \tilde{Z}(\Theta,\zeta)$ with $0\leq r \leq1$ and $0\leq z \leq1$.  The region inside the magnetic surface can be viewed as consisting of nested surfaces, which are generally not magnetic surfaces, of radius $\tilde{\rho} \equiv \sqrt{r^2 \tilde{R}^2 + z^2 \tilde{Z}^2}$.  When $\Theta$ and $\zeta$ are held fixed while differentiating $\rho$, the toroidal magnetic flux enclosed by the surface is then
\begin{eqnarray}
&&\psi = \int_0^1 dr \int_0^1 dz \int_0^{2\pi}d\Theta \vec{B}(R_a,Z_a)\cdot\hat{\zeta} \tilde{\rho} d\tilde{\rho}, \hspace{0.2in}\\
&& \mbox{  where  } \tilde{\rho} d\tilde{\rho} = \tilde{R}^2 r dr +\tilde{Z}^2 z dz. 
\end{eqnarray}

Once a number of field lines are followed, the $R(\Theta,\zeta)$, $Z(\Theta,\zeta)$, and $\psi$ for each is known.  The results for all these field lines defines a function $\vec{X}(\psi,\Theta,\varphi)$ and therefore $(\psi,\Theta,\varphi)$ magnetic coordinates for each $\psi$ studied.

%======================================================

\subsection{The $\psi$ derivative of $\vec{X}(\psi,\Theta,\zeta)$}

Sufficiently close to an arbitrary surface, a $(\rho,\Theta,\zeta)$ coordinate system can be defined using the unit normal of Equation (\ref{unit norm}).  Positions in $(R,\zeta,Z)$ coordinates are given by the function
\begin{eqnarray}
\vec{\mathcal{X}}(\rho,\Theta,\zeta) = \vec{X}_s(\Theta,\zeta) + \rho \hat{n}(\Theta,\zeta).
\end{eqnarray}
Sufficiently close means $\rho$ sufficiently small.

The toroidal magnetic flux in an infinitesimally thin annulus $d\rho$ is 
\begin{eqnarray}
\frac{d\psi}{d\rho}&=& \oint \vec{B}\cdot \left(\frac{\partial \vec{\mathcal{X}}}{\partial\rho}\times \frac{\partial \vec{\mathcal{X}}}{\partial\Theta} \right)d\Theta \\
&=& \oint \vec{B}\cdot \frac{ \left| \frac{\partial \vec{X}_s}{\partial\Theta} \right|^2  \frac{\partial \vec{X}_s}{\partial\zeta} - \frac{\partial \vec{X}_s}{\partial\Theta}\cdot\frac{\partial \vec{X}_s}{\partial\zeta} \frac{\partial \vec{X}_s}{\partial\Theta} }{\left| \frac{\partial \vec{X}_s}{\partial\Theta} \times \frac{\partial \vec{X}_s}{\partial\zeta} \right| } d\Theta. \
\end{eqnarray}
using the relation $\partial \vec{\mathcal{X}}/\partial\rho =\hat{n}$ and vector identities.  Since $\vec{X}(\psi,\Theta,\zeta)=\vec{\mathcal{X}}(\rho,\Theta,\zeta)$,
\begin{eqnarray}
\left(\frac{\partial \vec{X}}{\partial \psi}\right)_{\Theta\zeta} &=& \left(\frac{\partial \rho}{\partial \psi}\right)_{\Theta\zeta} \hat{n}\\
&=& \frac{\hat{n}(\Theta,\zeta)}{d\psi/d\rho}
\end{eqnarray}
is the $\psi$ derivative of the magnetic coordinates $\vec{X}(\psi,\Theta,\zeta)$ at fixed $\Theta$ and $\zeta$.
%===============================================

\subsection{The magnetic field line Hamiltonian}

A complete Hamiltonian description of the magnetic field lines $\Psi_p(\Psi,\Theta,\zeta)$ based on the $\vec{X}(\psi,\Theta,\zeta)$ coordinates can be obtained using the method given in 2023 by Boozer \cite{Boozer-H:2023}.  

Whether magnetic surfaces exist or not the exact magnetic field can be written in terms of two fluxes, $\Psi$ a toroidal flux and $\Psi_p$ a poloidal flux \cite{Boozer:RMP},
\begin{equation}
2\pi\vec{B}=\vec{\nabla}\Psi(\psi,\Theta,\varphi)\times\vec{ \nabla}\Theta + \vec{\nabla}\varphi\times\vec{\nabla}\Psi_p(\psi,\Theta,\varphi), \hspace{0.2in}
\end{equation}

The theory of general coordinates, see the two-page appendix to Reference \cite{Boozer:RMP}, implies
\begin{eqnarray}
\frac{\partial\Psi}{\partial\psi} &=&  2\pi \vec{B}\cdot\left(\frac{\partial \vec{X}}{\partial\psi}\times \frac{\partial \vec{X}}{\partial\Theta}\right) \\
&\equiv&F_\Psi(\psi,\Theta,\zeta) \hspace{0.2in} \mbox{   and   } \label{Psi-rho}\\
\frac{\partial\Psi_p}{\partial\psi} &=& 2\pi \vec{B}\cdot\left(\frac{\partial \vec{X}}{\partial\varphi}\times \frac{\partial \vec{X}}{\partial\psi}\right)\\
&\equiv& F_p(\psi,\Theta,\zeta) \label{Psi_p-rho}
\end{eqnarray}

The functions $F_\Psi(\psi,\Theta,\zeta)$ and $F_p(\psi,\Theta,\zeta)$ are only known at $\psi$ values associated with particular field line integrations.  However, they are analytic functions when the $\vec{X}_s(\Theta,\zeta)$ upon which they are based are analytic.  The implication is coefficients in the Fourier expansions $F_\Psi$ and $F_p$ have the forms $\psi^{m/2}$ times functions that can be Taylor expanded in $\psi$, so sensible interpolations in $\psi$ should be possible.

 By picking a regular array of $\Theta$ and $\zeta$ points, an ordinary differential equation,
\begin{equation}
\frac{d\psi}{d\Psi} = \frac{1}{F_\Psi(\psi,\Theta,\zeta)} \label{d rho / d Psi},
\end{equation}
can be integrated for each $(\Theta,\zeta)$ point to obtain $\psi(\Psi,\Theta,\zeta)$.  A Fast Fourier Transform then determines the coefficients in the Fourier series, which has only cosinusoidal terms when the field is stellarator symmetric;
\begin{equation}
\psi(\Psi,\Theta,\varphi) = \sum_{mn} \psi_{mn}(\Psi) \cos(m\Theta-n\zeta).
\end{equation}
For an analytic magnetic field, the Fourier amplitudes converge exponentially.

Similarly Equations (\ref{Psi_p-rho}) and (\ref{Psi-rho}) imply that
\begin{equation}
\frac{d\Psi_p}{d\Psi} = \frac{F_p(\psi,\Theta,\zeta)}{F_\Psi(\psi,\Theta,\zeta)}, 
\end{equation}
which can be integrated for each $(\Theta,\zeta)$ point from $\Psi_p(0)$ at $\Psi=\Psi_0$, while simultaneously integrating Equation (\ref{d rho / d Psi}),  to obtain $\Psi_p(\Psi,\Theta,\zeta)$.  Another Fast Fourier Transform then determines the coefficients in a Fourier series, which has only cosinusoidal terms when the field is stellarator symmetric;
\begin{equation}
\Psi_p(\Psi,\Theta,\zeta) = \sum_{mn} \Psi^p_{mn}(\Psi) \cos(m\Theta-n\zeta). \label{orig-B-line-Hamiltonian}
\end{equation}

$\Psi_p(\Psi,\Theta,\zeta)$ of Equation (\ref{orig-B-line-Hamiltonian}) is the Hamiltonian for the magnetic field lines expressed as a function of its canonical coordinates;
\begin{eqnarray}
&&\frac{d\Psi}{d\zeta} = - \frac{\partial \Psi_p}{\partial\Theta};\\
&&\frac{d\Theta}{d\zeta} = \frac{\partial \Psi_p}{\partial\Psi}.
\end{eqnarray}

The equations $\vec{X}(\psi,\Theta,\zeta)$ and $\psi(\Psi,\Theta,\zeta)$ give the spatial location of each point of the canonical coordinates $(\Psi,\Theta,\varphi)$.

%%%%%%%%%%%%%%%

%Actual magnetic fields are analytic functions of positions.  Any Fourier series with a finite number of terms is an analytic function even when it approximates the shape of a magnetic surface that is itself not analytic.  The Hamiltonian $\Psi_p(\Psi,\Theta,\zeta)$ obtained by the outlined procedure will be analytic, which implies the Fourier coefficients can be written as $\Psi^{m/2}$ times a Taylor expansion in $\Psi$.  This general form of the Fourier coefficients can be used to sensibly smooth and extrapolate the function $\Psi_p(\Psi,\Theta,\zeta)$.

%==============================

%%%%%%%%%%%%%%%%%%%%%%%%%%%%%%%%%%%%%%%%%%%%%

\section{Field-line chaos in cylindrical coordinates \label{chaos-cyl} }

Boozer and Elder \cite{Boozer-Elder:2021} gave a direct method for determining the extent to which a magnetic field line is chaotic and where in space the chaos occurs.  Their method is given here for consistency in notation and motivation.

Magnetic field lines are defined by $d\vec{X}/d\tau =\vec{B}(\vec{X})$.  The mathematics is simplest to take $\tau$ as just a parameter, but its physical interpretation is $d\tau = d\ell/B$, where $\ell$ the distance along a magnetic field line.  Using $(R,\zeta,Z)$ coordinates, $\vec{B}(\vec{X})=d\vec{X}/d\tau$ is
\begin{eqnarray}
\vec{B}(R,\zeta,Z)&=& \frac{\partial\vec{X}}{\partial R}\frac{dR}{d\tau} + \frac{\partial\vec{X}}{\partial \zeta}\frac{d\zeta}{d\tau} + \frac{\partial\vec{X}}{\partial Z}\frac{dZ}{d\tau} \\
&=& \frac{dR}{d\tau} \hat{R} + \frac{R}{N_p} \frac{d\zeta}{d\tau}\hat{\zeta} + \frac{dZ}{d\tau} \hat{Z}; \\
\frac{dR}{d\tau} &=& \vec{B}\cdot\hat{R}; \\
\frac{d\zeta}{d\tau} &=& \vec{B}\cdot\vec{\nabla} \zeta; \\
\frac{dZ}{d\tau} &=& \vec{B}\cdot\hat{Z}.
\end{eqnarray}
These equations can be integrated to obtain the magnetic field lines.  

Since $\vec{B}\cdot\vec{\nabla} \zeta$ is non-zero in the full region of interest in tokamaks and stellarators, one can obtain field lines $\vec{x}(\zeta)$ in  $(R,\zeta,Z)$ coordinates by integrating
\begin{eqnarray}
\frac{dR}{d\zeta} &=& \vec{\mathcal{R}}\cdot\hat{R}; \\
\frac{dZ}{d\zeta} &=& \vec{\mathcal{R}}\cdot\hat{Z},  \hspace{0.2in}\mbox{where    }\\
\vec{\mathcal{R}}(R,\zeta,Z) & \equiv& \frac{\vec{B}}{ \vec{B}\cdot\vec{\nabla} \zeta} \\
&=& \mathcal{R}_R \hat{R} + \hat{\zeta} + \mathcal{R}_Z \hat{Z}.
\end{eqnarray}
These equations give the locations of the field lines in constant $\zeta$ planes.  The $R$ and the $Z$ have the properties of ordinary Cartesian coordinates in those planes with the positions of the lines $\vec{X}(\zeta)=R(\zeta)\hat{R}(\zeta) +Z(\zeta)\hat{Z}(\zeta)$.

To study magnetic flux tubes, the behavior of all field lines infinitesimally separated from an arbitrarily chosen line $\vec{X}_0(\zeta)$ is needed.  Let  $\vec{\delta}=\delta_R\hat{R}+\delta_Z\hat{Z}$ be the infinitesimal separation between a line and the arbitrarily chosen line $\vec{X}_0(\zeta)$ in the $(R,Z)$ planes of $\zeta$.  Then,
\begin{eqnarray}
\frac{d (\vec{X}_0 +\vec{\delta})}{d\zeta} &=& \vec{\mathcal{R}}(\vec{X}_0+\vec{\delta}),  \mbox{   so   }\\
\frac{d \vec{\delta}}{d\zeta} &=&\vec{\mathcal{R}}(\vec{X}_0+\vec{\delta}) - \vec{\mathcal{R}}(\vec{X}_0);\\
&=& (\vec{\delta}\cdot\vec{\nabla})\vec{\mathcal{R}} \hspace{0.1in}\mbox{as}\hspace{0.1in} |\vec{\delta}|\rightarrow0.
\end{eqnarray}

The two components of $\vec{\delta}$ in the constant $\zeta$ planes are related to the two components of $\vec{\mathcal{R}}$  in those planes by
\begin{eqnarray}
\frac{d\delta_R}{d\zeta} &=& \delta_R \frac{\partial \mathcal{R}_R}{\partial R} + \delta_Z\frac{\partial \mathcal{R}_R}{\partial Z}  \label{ delta-R}\\
\frac{d\delta_Z}{d\zeta} &=& \delta_R \frac{\partial \mathcal{R}_Z}{\partial R} + \delta_Z\frac{\partial \mathcal{R}_Z}{\partial Z}.  \label{ delta-Z}
\end{eqnarray}
%Note the derivative $\delta_R(d\hat{R}/d\zeta)= \delta_R \hat{\zeta}/N_p$ but $\vec{\mathcal{R}}\cdot\hat{\zeta}=1$.

The coupled ordinary differential equations for $\delta_R$ and $\delta_Z$, Equations (\ref{ delta-R} and \ref{ delta-Z}), should be integrated from an initial $\zeta$ with two different initial conditions: (i) $\delta_R^{(1)}\equiv\delta_{RR}=1$ and $\delta_Z^{(1)}\equiv\delta_{RZ}=0$ and (ii) $\delta_R^{(2)}\equiv\delta_{ZR}=0$ and $\delta_Z^{(2)}\equiv\delta_{ZZ}=1$.  Since the equations are linear, the initial size of the $\delta$'s is irrelevant.

The Frobenius norm of a matrix is the square root of the sum of the squares of the matrix elements and is also equal to the square root of the sum of the squares of the singular values of a singular value decomposition (SVD) of the matrix.   The Frobenius norm of the matrix of the separations measures the stretching of an initial circular flux tube of magnetic field lines into an ellipse,
\begin{equation}
e^{\sigma_c(\zeta)} \equiv  \sqrt{\delta_{RR}^2 + \delta_{RZ}^2 + \delta_{ZR}^2 + \delta_{ZZ}^2}.
\end{equation}
The parameter  $\sigma_c(\zeta)$ measures the degree of chaos and where along the line the degree of chaos rapidly increases.   A region in which $\sigma_c>>1$ is highly chaotic. 

%%%%%%%%%%%%%%%%%%%%%%%%%%%%%%%%%%%%%%%%%%%%%%%%%%

\section{Importance for Divertor and Disruption Calculations \label{Importance}}

An efficient method for studying the transition in space from true magnetic surfaces to leaky surfaces, where the leaks define collimated tubes of magnetic flux, is required for studying divertors, particularly non-resonant divertors, Section \ref{Divertors}.  The transition in both space and time is needed to understand the radial spread of heat and in the net plasma current, more precisely $j_{||}/B$, during a disruption, Section \ref{Disruptions and runaway}.  But for disruption studies, it is even more important to determine the formation of runaway electrons beams in collimated tubes of magnetic flux from leaky surfaces that produce a sufficient concentration in space and time to burn holes through the chamber walls, Section \ref{Disruptions and runaway}.

\subsection{Divertors \label{Divertors} }

The primary role of a divertor is now generally viewed as a way to channel the plasma particles in the exhaust into highly localized pumps.  The pump area is limited to of order a percent of the wall area by the need to have almost all of the wall thin to allow sufficient tritium breeding.  To have even a perfect-pump work, which is a black-hole for neutral particles, the neutral density must be approximately two orders of magnitude larger at the pump entrance than the average neutral density in the region between the plasma and the chamber wall.  This is called the compression ratio.  Otherwise, the plasma will ionize the impinging neutrals, and a large circulation of neutrals between the plasma and the wall will occur, which erodes the wall.  

%It may be possible to cover the wall with a thin layer of liquid metal, such as lithium, that could capture the exhaust deuterium, tritium, and helium with the flow of the liquid metal carrying the exhaust to a highly localized removal location \cite{Lithium:2023}.  If this is feasible, divertor design could be revolutionized.

In the past, it was hoped that magnetic field line chaos outside the core plasma could spread the power exhaust over the walls avoiding hot spots. Viana et al  \cite{Chaotic divertor:2011} studied the mathematical properties of a chaotic edge and found that the avoidance of hot spots is extremely difficult.  

%The thermal power deposition on the walls must be close to uniform.  The higher the power density in Watts per square meter, the lower is the cost of fusion power in dollars per Watt until the limits of materials are reached at $\sim~$10 MW/m$^2$.  The cost of fusion power is essentially given by the capital cost of the machine which is approximately proportional to the area of the walls of the plasma chamber.  The thickness of the surrounding chamber, $\sim 1.5~$m, is determined by nuclear physics: tritium breeding and shielding the coils from neutrons.

High-Z radiation from argon or neon impurities at the plasma edge can cool the plasma from a central temperature of about 10~keV to a thermal power density sufficiently low and uniform that it can be tolerated by the walls.  The use of high-Z impurities places a strong constraint on divertor design.  The high-Z impurities must have a sufficient density to radiate the out-coming thermal power, but only in the edge, not in the regions where the fusion power is being generated.  A suggestion for accomplishing this is given in \cite{Boozer;2024NF}.

The location of the divertor on the chamber wall is determined similarly for tokamaks and stellarators.  The diverted plasma strikes the wall where magnetic field lines that penetrate the wall come sufficiently close to the confined plasma for cross-field diffusion to load plasma on the lines before the plasma flows to the wall along the lines.  

%The normal magnetic field to the wall is non-zero over most of the wall.  $\vec{\nabla}\cdot\vec{B}=0$ implies that each magnetic field line that enters the plasma chamber must exit the chamber by striking the wall.  The number of toroidal periods a field line transits between entry and exit tends to infinity for the field lines that pass closest to a magnetic surface.  This is illustrated by calculations done by Kelley Garcia and collaborators \cite{Garcia:2024}.  In tokamaks a typical field line that enters through the wall makes several toroidal transits before exiting through the wall.  The helical field in a stellarator produces a rotational transform in quadratic order, which makes the normal field far stronger in a stellarator than in a tokamak with a similar rotational transform.  A typical field line in a stelllarator goes through only a faction of a period between entering and leaving the chamber.   The field lines that come sufficiently close to the main plasma body to be loaded with plasma form only narrow helical stripes on the wall.  These helical stripes must have equal magnetic flux at entry and exit.  Plasma strikes the walls flowing both parallel and antiparallel to the magnetic field lines.

%%%%%%%%%%

Four questions are important in the design of divertors:  (1) Where do the tubes of magnetic flux that carry plasma strike the chamber walls? (2) How much magnetic flux is in these tubes versus the distance of closest approach of the lines to the outermost magnetic surface?  (3) How much magnetic flux lies on open magnetic field lines just outside the outermost magnetic surface that have a sufficiently short connection length that the flow time to the divertor chambers is short compared to the cross-field diffusion time into the plasma.  This allows (a) high-Z impurities to be sown in the edge and radiate the power without diffusing into the plasma core and (b) recirculating neutrals to be ionized and carried to the divertor rather than charge exchanging within the hot plasma.   The production of hot neutrals by charge exchange would erode the walls and could make the neutral compression ratio too low for the divertor pumps to provide adequate particle removal.  (4) Over what range of machine operations can the answers to the first three questions be controlled?   Issues number two, the magnetic flux in stripes on the wall, and three, the flux in region near the plasma,  are related.  The difference is how many transits of a period a field line lingers near the plasma versus the number of transits it makes between leaving the vicinity of the plasma and striking the wall.

The design of the divertor for the European tokamak DEMO \cite{DEMO-divertor:2022} illustrates issues with an axisymmetric single-null divertor and methods that have been devised to address them.  As demonstrated by J.-K. Park and collaborators \cite{Park:2018}, external coils can be used to make the plasma sufficiently non-axisymmetric to stabilize ELMs while preserving quasi-axisymmetry in the core, so the overall plasma confinement is that of an axisymmetric tokamak.  An implication is that non-axisymmetric divertor designs are relevant to tokamaks as well as to stellarators.

Poincar\'e tangles, Appendix \ref{Tangle}, imply the outermost magnetic surface in a tokamak can become very rough with a small asymmetry prone to forming a chaotic layer.  Indeed, the Poincar\'e plots in \cite{DIII-D tangle,Punjabi-tangle:2014,Stickiness:2014} all show regions of chaos as well as the tangles in the edge region.   Chaos implies the forward and backwards separatrix surfaces are not perfect; they are  leaky barriers especially in the vicinity of the X-line.  A larger asymmetry generally gives greater chaos.

The properties of stellarator divertors, in particular the distinction between resonant and non-resonant divertors, was discussed Section 3 of a 2015 paper on stellarator design by Boozer \cite{Boozer-Stell-design:2015}.  A resonant divertor requires a particular value of the rotational transform at the plasma edge; a non-resonant divertor does not.

The most studied divertor concept for stellarators is a resonant divertor, the island divertor used on the W7-X device.  The island divertor concept was given 1996 by Erika Strumberger \cite{Strumberger:1996}, and its successful operation was a part of a 2024 review of the first Wendelstein 7-X long-pulse campaign  \cite{W7-X_div2024}.

Earlier, in 1992, Erika Strumberger \cite{Strumbeger:divertor1992} pointed out that designs for W7-X had a naturally arising non-resonant divertor. N\"uhrenberg and Strumberger published more detailed calculations for this type of divertor \cite{Nuhrenberg-Strumberger:1992}.

The naturally-arising status of the non-resonant divertor raised concerns about its robustness.  Aaron Bader and collaborators have found that the non-resonant divertor is remarkably robust \cite{Bader-Boozer:2017,HSX divertor:2018} and that this robustness is due to divertor action being localized near places of strong surface curvature \cite{HSX divertor:2018}.  The importance of local surface curvature was implied, but not explicitly stated, in the discussion of ``helical edges" by N\"uhrenberg and Strumberger \cite{Nuhrenberg-Strumberger:1992}.

Without using the more efficient method outlined in this paper, Punjabi and Boozer \cite{Boozer-Punjabi:2018,Punjabi-Boozer:2022}, as well as by Garcia et al \cite{Garcia:2024}, have studied the width of the flux tubes that strike the wall versus the distance of closest approach of the tubes to the outermost confining magnetic surface of the plasma.  The presence of large islands outside the outermost confining surface was found by Garcia et al to fundamentally change the relationship between the flux tube width and how closely the tubes come to that outermost surface.  But, the presence of large islands does not destroy the robustness of where field lines strike the walls.  An exception to this robustness occurs when the islands themselves strike the walls, as in the W7-X divertor.

%In 2018, Boozer and Punjabi \cite{Boozer-Punjabi:2018} derived a simple Hamiltonian that gave an efficient method for determining the probability of a trajectory being lost in a toroidal transit versus its distance in flux, $\psi-\psi_b$, outside the outermost confining magnetic surface.  They found the probability obeys a power law, approximately $(\psi-\psi_b)^2$.  In further studies, Punjabi and Boozer \cite{Punjabi-Boozer:2022} found the power in the power law depends on the structure of the field but is comparable to two.  Kelly Garcia and coauthors \cite{Garcia:2024} have studied the relation among the length of field lines $L_c$ that strike the walls, their closest approach $\delta_N$ to the outermost confining plasma surface for various configurations that could be produced by the HSX coils.  When islands are small, a power-law relationship is found $L_c \propto 1/\delta_N^{1.7}$.  The 1.7 power is analogous to similar powers found by Punjabi and Boozer \cite{Boozer-Punjabi:2018,Punjabi-Boozer:2022}.  The presence of large islands outside the outermost confining surface  was found by Garcia et al \cite{Garcia:2024} to fundamentally change this dependence but not the robustness of where field lines strike the walls.  An exception to this robustness occurs when the islands themselves strike the walls, as in the W7-X divertor.

%%%%%%%%%%%%%%%%%%%%

\subsection{Disruptions and runway electrons \label{Disruptions and runaway} }

Nicholas Eidietis \cite{Eiditis:2021} has emphasized that fusion power based on tokamaks cannot be feasible until methods of avoiding disruptions are developed.  Disruptions are associated with the sudden breakup of the magnetic surfaces---approximately six orders of magnitude faster than would be expected from the resistive diffusion of magnetic flux---and rapid plasma cooling due to the sweeping of impurities into the plasma core.  The physics of these processes involves the physics of chaos such as those described by Contopoulos and Harsoula \cite{Stickiness:2010} and is given in more detail in a 2024 article by Boozer \cite{Disruption E-field:2024} on the electric field in a disrupting plasma.  

Disruptions pose a far greater obstacle to fusion power using tokamaks than stellarators.  Nevertheless, all toroidal fusion concepts are subject to the breakup of the magnetic surfaces on a timescale set by an ideal instability that does not self-saturate as shown  by Boozer \cite{Boozer-rapid:2022} in 2022.  Simulations performed in the same year by Jardin et al \cite{Jardin-rapid:2022} were consistent with an ideal instability explaining extremely fast disruptions in  NSTX.  The small-scale structures associated with chaos that give rise to this effect are challenging to convincingly resolve numerically but must be understood for reliable predictions of the performance of power-plant scale fusion devices.

Breizman, Aleynikov, Hollmann, and  Lehnen \cite{Runaway:2019} have reviewed the physics of what may be the most intractable feature of tokamak disruptions, the transfer of the plasma current from being carried by near-thermal to $\sim 10~$MeV electrons by a runaway avalanche.  The runaway electrons have only about ten percent of the total energy of the pre-disruption thermal plasma.  However, the thermal energy is generally lost by impurity radiation, which is spread quasi-uniformly over the walls, while the relativistic electrons can strike the walls in a way that is highly localized in space and time, which causes unacceptable damage.   

Hope that a solution may exist to this issue was raised with the 2021 publication of experiments on JET \cite{Reux:2021} and DIII-D  \cite{Soldan:2021} that showed a strong instability removed a large current carried by relativistic electrons with no noticeable  power concentration on the walls.   The physics behind this result had been described five years earlier by Punjabi and Boozer \cite{Boozer-Punjabi:2016}.  When relativistic electrons are in a large region of chaotic magnetic field lines but confined by an annulus of magnetic surfaces, the speed of breaking the surfaces in this annulus determines whether the energy in these electron is concentrated or spread over the walls.  When the timescale of the annulus destruction is slow compared to the time for a relativistic electron to cover the chaotic region, then the energy deposition will be highly concentrated due to effects such as those described in Viana et al  \cite{Chaotic divertor:2011} and the Altmann et al review \cite{Leaky systems:2013}.  When the annulus destruction is far more rapid than this relativistic electron coverage time, the relativistic electrons strike the wall quasi-uniformly.

%%%%%%%%%%%%%%%%%%%%%%%%%%%%%%%%%%%%%%%%%%%%%%%%%%%%%%%
\section*{Acknowledgements}

This material is based upon work supported by the Grant 601958 within the Simons Foundation collaboration Hidden Symmetries and Fusion Energy, and by the U.S. Department of Energy, Office of Science under Award Nos. DE-FG02-95ER54333, DE-SC0024548, and DE-AC02-09CH11466. % The SC grant is HifiStell; the AC grant is Stellfoundry
 \vspace{0.01in}

\section*{Author Declarations}

The author has no conflicts to disclose. \vspace{0.01in}

%%%%%%%%%%%%%%%%%%%%%%%%%%%%%%%%%%%%%%%%%%%%%%%%%%%%%%%%%%%%%%%%

\section*{Data availability statement}

Data sharing is not applicable to this article as no new data were created or analyzed in this study.

%%%%%%%%%%%%%%%%%%%%%%%%%%%%%%%%%%%%%%%

\appendix

%=================================================

\section{Discrete Evaluation of Fourier Integral \label{Discrete}}

Equation (\ref{fourier}) for $Z_f(\omega)$ can be evaluated using a sum.  Assume $Z(\zeta)$ is known at discrete points in $\zeta$ with $\zeta_d$ the distance between these points, so $\zeta = \zeta_d j$ with $j\geq0$ an integer.   Equation (\ref{fourier}) has the discrete form
\begin{eqnarray}
Z_f(\omega) &=& 2\zeta_d \sum_{j=0}^\infty G(\zeta_d j) \cos(\omega\zeta_d j) Z(\zeta_d j)\\
&=& 2\zeta_d \Big(\sum_{j=0}^\infty \frac{e^{-\frac{1}{2}\left(\frac{\zeta_d j}{\lambda}\right)^2}}{\lambda\sqrt{\pi/2}}  \cos(\omega\zeta_d j) Z(\zeta_d j)\nonumber\\&& \hspace{0.2in} -\frac{Z(0)}{2\lambda\sqrt{\pi/2}}\Big)
\end{eqnarray}
The final term subtracts half of the contribution of the $j=0$ term, which must be done to obtain unity for the sum over the Gaussian window function.

Equation (\ref{transform}) is the equation that must be accurately calculated.  Let $\lambda_d \equiv \lambda/\zeta_d$ and $\hat{\omega}\equiv (\omega-\omega_{mn})\zeta_d$, then this equation has the form and the supposed equality
\begin{eqnarray}
 \frac{\sum_{j=0}^{\infty}e^{-\frac{1}{2}\left(\frac{j}{\lambda_d}\right)^2}   \cos(\hat{\omega} j)  - \frac{1}{2}}{\lambda_d \sqrt{\pi/2}} = e^{-\frac{\lambda_d^2 \hat{\omega}^2}{2} },
\end{eqnarray}
which can be easily checked and works when $\hat{\omega} \lambda_d \gtrsim 5$.  Only a limited number of $j$'s contribute significantly to the sum, $j \lambda_d \lesssim 5$, so only the $j \lesssim 25/\hat{\omega}$ are relevant.

%%%%%%%%%%%%%%%%%%%%%%%%%%%%%%%%%%%%%%%

\section{Poincar\'e tangles \label{Tangle}}

Small non-axisymmetric perturbations to an axisymmetric single-null divertor produce what is called a Poincar\'e tangle because it arose in his 1890 proof \cite{Poincare:1890} that the three-body gravitational problem does not have an analytic solution.  A Poincar\'e tangle is a set of lobe-like structures on a Poincar\'e plot that have an amplitude that increases without limit the closer the lobe is to the X-point. The X-point, which is a closed magnetic field line and should be called an X-line, is moved but not eliminated by a small perturbation.   

When the X-line is enclosed by a  circular tube of infinitesimal radius,  four curves on that tube are of special importance.  In axisymmetry, field lines that intercept two of these curves define the two legs of the divertor---plasma flows parallel to the magnetic field lines from one and antiparallel from the other to a divertor chamber.  For the discussion of tangles, the more important pair of curves on the tube enclosing the X-line are intercepted by the field lines that lie on the separatrix.  The separatrix separates the region in which field lines lie on nested magnetic surfaces that enclose the plasma from open magnetic field lines that can be followed from where they enter to where they leave the plasma chamber.  

In axisymmetry the same separatrix surface is obtained whether field lines are integrated forwards or backwards from the enclosing tube about the X-line.  However, these two ways to define the separatrix surface differ in the presence of asymmetric perturbations.  Simple geometry implies that a perturbation of given amplitude has a greater effect on the path followed by a field line the greater the distance of the perturbation from the X-line.  This was proven  in 1981 by Boozer and Rechester  \cite{Boozer-Rechester:1981}.   When forwards field line integrations are used to define a surface, that surface starts out smooth.  When the lines are followed all the way around until they strike the enclosing tube, the surface picks up oscillations which oscillate ever faster in the poloidal plane as they approach the X-line.  This comes from the poloidal magnetic field vanishing at the X-line, so each toroidal circuit advances a line a smaller distance poloidally the closer the X-line.   When backward field line integrations are used to define the surface, the surface again starts out smooth but ends up with large oscillations as the X-line is approached from the other direction.  These two surfaces serve as barriers for field lines and, therefore, define a flux tube.  The poloidal width of the oscillations determines the width of the flux tube.  As that width becomes narrower, the magnitude of the oscillations, called lobes, must increase in order to conserve the enclosed toroidal flux.  The amplitudes of the lobes go to infinity as the X-line is approached from either side.   The complexity of the resulting field line trajectories---especially when projected on a plane defined by constant toroidal angle---led to the term tangle.   

In 2009, Wingen, Evans, and Spatschek  \cite{DIII-D tangle} discussed the effect of the tangles on the DIII-D divertor.   In 2014, Punjabi and Boozer \cite{Punjabi-tangle:2014} calculated the tangle expected in tokamak divertors and illustrated their results with a number of figures.  The properties of strongly perturbed single-null axisymetric divertor were studied in a 2014 paper by Martins, Roberto, and Caldas \cite{Stickiness:2014}.   In 2018, Boozer and Punjabi \cite{Boozer-Punjabi:2018} found that a quadrupole, $\sin(2\theta-\varphi)$, perturbation to the poloidal magnetic flux gave solutions with fundamentally different characteristics depending on whether the perturbation was larger or smaller than $2\times10^{-4}$.

%===========================================================================

\end{document}